\documentclass[%
 reprint,
 jmp,
superscriptaddress,
 amsmath,amssymb,
 aps,
]{revtex4-2}

\usepackage{graphicx}
\usepackage{dcolumn}
\usepackage{bm}
\usepackage{mathtools}

\usepackage{braket}
\usepackage{xcolor}
\begin{document}

\preprint{APS/123-QED}

\title{Photon number distribution of squeezed light from a silicon nitride microresonator measured without photon number resolving detectors}

\author{Emanuele Brusaschi}
\author{Massimo Borghi}
\email{corresponding author: massimo.borghi@unipv.it} 
\author{Marcello Bacchi}
\affiliation{Dipartimento di Fisica, Università di Pavia, Via Bassi 6, 27100 Pavia, Italy.}
\author{Marco Liscidini}
\author{Matteo Galli}
\affiliation{Dipartimento di Fisica, Università di Pavia, Via Bassi 6, 27100 Pavia, Italy.}
\author{Daniele Bajoni}
\affiliation{Dipartimento di Ingegneria Industriale e dell'Informazione, Università di Pavia, Via Ferrata 5, 27100 Pavia, Italy.}


\date{\today}

\begin{abstract}
The measurement of the photon number distribution (PND) allows one to extract metrics of non-classicality of fundamental and technological relevance, but in principle it requires the use of detectors with photon number resolving (PNR) capabilities. 

In this work we reconstruct the PND of two-mode pulsed squeezed light generated from a silicon nitride microresonator using threshold detectors and variable optical attenuations. The PNDs are characterized up to $\sim 1.2$ photons/pulse, through which we extracted an on-chip squeezing level of $6.2(2)$ dB and a noise reduction factor of $-3.8(2)$ dB. The PNDs are successfully reconstructed up to an Hilbert space dimension of $6\times6$. The analysis performed on the photon-number basis allows us to characterize the influence of a spurious thermal background field that spoils the photon number correlations. We evaluate the impact of self and cross phase modulation on the generation efficiency in case of a pulsed pump, and validate the results through numerical simulations of the master equation of the system.
\end{abstract}

\maketitle


\section{\label{sec:intro} Introduction}
The characterization of quantum optical states in the photon number basis is of great relevance in fundamental science, quantum communication, and quantum information processing. For example, non-classical optical states are characterized by a negative Mandel-Q parameter, measuring the departure of their PND from that of a coherent state \cite{loudon2000quantum}. Similarly, from the PND we can determine the sign of the Wigner function at the origin \cite{harder2014time} and of the eigenvalues of the matrix containing the high-order correlation moments \cite{laiho2022measuring}. In both cases, a negative value represents a criterion of non-classicality. In decoy-state quantum key distribution (QKD), one can estimate the amount of information
an eavesdropper could get by sending pulses with different mean photon numbers and of which the PND is known \cite{shan2021characterizing}. The latter requirement stems from the fact that even if mean photon number is known, the photon number
distribution still affects the performance of the QKD system. Therefore, it is important to characterize it to select the states which are more robust to photon number splitting attacks \cite{dynes2018testing}. \\
The PND of bipartite optical states carries information on both inter-beam photon number correlations and intra-beam photon statistics.
Each of the conjugate beams generated by spontaneous parametric down conversion (SPDC) or spontaneous four wave mixing (SFWM), is usually described by a multi-mode thermal state. The number of these modes is called the Schmidt number and determines the amount of entanglement of the bipartite state \cite{christ2011probing}. The Schmidt number is tailored on the requirements of the application. For example, in quantum computing and simulation, single photons should be heralded in a well defined spectro-temporal mode to achieve high visibility quantum interference \cite{shchesnovich2014sufficient}.
Inter-beam correlations in the photon number are instead quantified by a reduced variance in the photon number difference between the two beams with respect to that of a coherent state \cite{vaidya2020broadband}. Finally, the shrinkage of the variance is called noise reduction factor (NRF). A value below unity indicates photon number difference squeezing, which is a signature of non-classicality. \\
A prominent example of a quantum optical state with strong photon number correlation is squeezed light, which is a fundamental resource for quantum information processing and simulation. In particular, squeezed light can be efficiently generated from integrated microresonators. Broadband single \cite{zhang2021squeezed} and two-mode \cite{vaidya2020broadband,zhao2020near} quadrature squeezing have been already reported for these types of sources and characterized by coherent photodetection. \textcolor{black}{While homodyne tomography allows us to fully reconstruct the complex Wigner function of the optical state, and thus also the coherences that may exist between different Fock states, its implementation can be very challenging for two-mode squeezed light because of the need of phase locking two local oscillators to the pump beam with hundreds of GHz of separation. \cite{vaidya2020broadband,zhang2021squeezed}}. 
Alternatively, the PND of squeezed light can be measured by photon number resolving detectors (PNR). For example, transition edge sensors \cite{schmidt2018photon} have been used in \cite{vaidya2020broadband} to measure the NRF of squeezed light generated by a silicon nitride resonator, and in \cite{arrazola2021quantum,madsen2022quantum} to detect patterns of bosonic coalescence at the output of sampling machines. However, PNR detectors have long dead-times ($>$ tens of $\mu s$), they are not commercial and they work at temperatures below 100 mK.\\
The large-scale deployment of integrated sources of squeezed light on different technology platforms and in multidisciplinary research fields drive the demand for characterization procedures which are faster, cheaper and simpler than those which implement PNR detectors. An ideal solution is  the use of standard superconducting nanowire single photon detectors (SNSPD) or off-the-shelf avalanche photodiodes. 
However, their PNR capability is so far limited to a small number of photons ($<3-4$) and they heavily rely on electronic post processing \cite{cahall2017multi}. 
Nevertheless, they can provide PNR capabilities when employed in temporal or spatial multiplexing schemes. These strategies fall under the umbrella of pseudo-PNR, which include fiber loop detectors \cite{banaszek2003photon}, beamsplitter trees \cite{piacentini2015positive}, cascaded delay-lines \cite{harder2014time} . \\
A parallel strategy for retrieving the PND that is efficient in hardware resources exploits the nonlinear photodection probability to different photon number states \textcolor{black}{\cite{Banner:24}} and in presence of losses \cite{mogilevtsev1999reconstruction}. The nonlinearity introduces photon number
sensitivity to the detected on-off statistics, from which the photon number distribution can be inferred by statistical
methods, such as the Expectation-Maximization algorithm (EM) \cite{vardi1993image}. The PND of a broad class of quantum optical states have been retrieved using this strategy, including heralded photons from SPDC, superposition of single photon states, coherent and thermal states \cite{zambra2005experimental}.
\\
In this work, we extend this approach  to pulsed two-mode squeezed vacuum (TMS) generated by SFWM in a silicon nitride microresonator. We show that from the reconstructed PND we can detect the presence of parasitic thermal noise, mainly originating from spontaneous Raman scattering in the resonator, which reduces the photon number correlation between the squeezed modes. 
We recover the average photon number per pulse, the squeezing parameter, and the NRF.  
Finally, we compare our results with numerical simulations of the master equation of the system, including the influence of self and cross-phase modulation (SPM/XPM), and discuss the limits of application of the method. 

\section{\label{sec:theory} Theory of photon number reconstruction using on-off detectors}
Here we briefly review the method presented in \cite{brida2011quantum} for the reconstruction of the PND using on-off detectors and variable attenuation. The PND $\rho_n$ of a single mode quantum optical state is given by \mbox{$\rho_n = \textrm{Tr}(\hat{\rho}\ket{n}\bra{n})=\left[ \hat{\rho}\right ]_{nn}$}, where  $\left[ \hat{\rho}\right ]_{nn}$ are the diagonal elements of the density matrix $\hat{\rho}$ on the photon number basis $\ket{n}$. 
Here we define \mbox{$\ket{n}=(n!)^{-\frac{1}{2}}\left (A^{\dagger}\right)^n\ket{\textrm{vac}}$}, where the operator $A^{\dagger}$ creates a single photon in a well defined spectral-temporal mode. The operator $A^{\dagger}$ will be associated in Section \ref{subsec:PND_squeezed} to the single Schmidt mode in which the signal and the idler photon will be generated.
The measurement process of an on/off detector with detection efficiency $\eta$ is described by a two-value positive operator-valued measure (POVM) $\Pi_{\textrm{on}/\textrm{off}}(\eta)$. The off-event in which no photons are detected is represented by
\begin{equation}
\Pi_{\textrm{off}} (\eta) = \sum_{n= 0}^{\infty} (1-\eta)^n\ket{n} \bra{n}, \label{eq:pi_off}
\end{equation}
and the related probability outcome  \mbox{$p_{0}(\eta) = \textrm{Tr}(\Pi_{\textrm{off}(\eta)}\hat{\rho})$} is
\begin{equation}
    p_{0}(\eta) = \sum_{n = 0}^{\infty} (1-\eta)^n \rho_n. 
\end{equation}
By performing $M$ measurements with different quantum efficiencies $\eta_\mu$ ($\mu = 1,...,M)$ we construct a vector $\boldsymbol{p_0}$ of off-events probabilities
\begin{equation}
    \boldsymbol{p_0} = \big[p_{0}(\eta_\mu) : p_{0}(\eta_\mu) = \sum_{n = 0}^{\infty} B_{\mu n} \rho_n \quad \mu = 1,2,..,M \big], \label{eq:p_0}
\end{equation}
where $B_{\mu n}=(1-\eta_\mu)^n$. Equation (\ref{eq:p_0})  corresponds to a linear system that can be written in compact form as $\boldsymbol{p_0} = B \bm{\rho}$.
It can be interpreted as a linear positive statistical model for the unknown parameters $\rho_n$ that can be solved through the EM algorithm \cite{brida2011quantum}. In this iterative approach, the estimate of the PND element $\rho_n^{(i+1)}$ at the step $i+1$ is calculated as
\begin{equation} \label{algoritmo_EM}
    \rho_n^{(i+1)} = \rho_n^{(i)} \sum_{\mu = 1}^M \frac{B_{\mu n} f_{0}(\eta_\mu)}{ (\sum_{\nu} B_{\nu n}) p_{0}(\eta_\mu) \{ \boldsymbol{\rho^{(i)}} \} }, 
\end{equation}
where $f_{0}(\eta_\mu)$ is the experimental off-probability measured at  the detection efficiency $\eta_{\mu}$, while $p_{0}(\eta_\mu) \{ \bm{\rho}^{(i)} \}$ is calculated by Eq.(\ref{eq:p_0}) through the estimate of $\bm{\rho}$ at step $i$.
The convergence of the algorithm is checked by evaluating the error parameter $\epsilon$, which is defined as the geometric distance between $f_0({\eta_{\mu}})$ and the reconstructed off-probabilities at the $i^{\textrm{th}}$ step
\begin{equation} \label{error_formula}
    \epsilon^{(i)} = \frac{1}{M} \sum_{\mu = 1}^{M} | f_{0}(\eta_\mu) - p_{0}(\eta_\mu) \{ \boldsymbol{\rho^{(i)}}  \} |.
\end{equation}
The similarity between $\bm{\rho}$ and a target theoretical distribution  $\boldsymbol{\rho}^{\textrm{theo}}$ is evaluated with the fidelity $\mathcal{F}$, defined as
\begin{equation}
    \mathcal{F}^{(i)} = \sum_{n=0}^{N} \sqrt{\rho_n^{(\textrm{theo})}\rho_n^{(i)}}.
\end{equation}
The method can be generalized to the bipartite case (see Appendix C) to reconstruct a two-dimensional PND $\rho_{nn'}$ from the on-off click probabilities and the coincidences between the two channels $x$ and $y$. In the bipartite case there are four possible outcomes $p_{xy}(\eta)$ for each measurement, where $x(y)=0$ labels a no-click event in the detector $x(y)$, while $x(y)=1$ denotes a click. \\
In general, the efficacy of the method depends on the maximum efficiency $\eta_{\textrm{max}}$ and on the level of noise in the data. First, if $n\eta_{\textrm{max}}\ll 1$, then \mbox{$(1-\eta_{\textrm{max}})^n\sim 1-\eta_{\textrm{max}}n+\mathcal{O}((n\eta_{\textrm{max}})^2)$}, and Eq.(\ref{eq:p_0}) reduces to \mbox{$p_{0}(\eta_\mu)=\sum_n \eta n\rho_n=\eta \langle n \rangle$}, i.e, the method can only solve the average number of photons but not the individual $\rho_n$. Hence, we require that \mbox{$\eta_{\textrm{max}}\sim\frac{1}{\langle n \rangle}$} for optimal performance, which means that losses should be kept as low as possible in order to reconstruct the PND of states with small $\langle n\rangle$. Second, if we apply the binomial expansion in Eq.(\ref{eq:p_0}) we obtain \mbox{$p_{0}(\eta) = \sum_{k=0}^{\infty} \left (\sum_{n=0}^k \rho_n C_{nk}\right )\eta^k=\sum_{k=0}^{\infty}Q_k\eta^k$}, where \mbox{$C_{nk} = (-1)^k{n\choose k}$} and $Q_k=\sum_{n=0}^{k}\rho_nC_{nk}$. Then, $\frac{\partial^{m}p_{0}}{\partial\eta^m}=m!Q_m$, which implies that only the elements of $\rho_n$ from $n=0$ to $n=m$ contribute to the $m^{\textup{th}}$ derivative of $p_{0}$. Thus, the next leading order term $\rho_{m+1}$ in the PND can be reliably estimated only if $\frac{\partial^{m+1}p_{0}}{\partial\eta^{m+1}}$ is accurately determined from the experiment, which imposes stringent demands on the precision and accuracy with which $p_{0}(\eta)$is measured, especially for large $n$. Rigorous Cramér-Rao bounds on the estimate of $\rho_n$ are derived in \cite{brida2011quantum}.
In Appendix B we experimentally validate the method by reconstructing the PND of known single-mode optical thermal and the coherent states. In both cases, we obtain fidelities higher than $0.99$ with the theoretical target distributions.

\subsection{Photon number statistics of squeezed light from a lossy
microresonator \label{subsec:PND_squeezed}}
\begin{figure}[b!]
    \includegraphics[scale = 0.83]{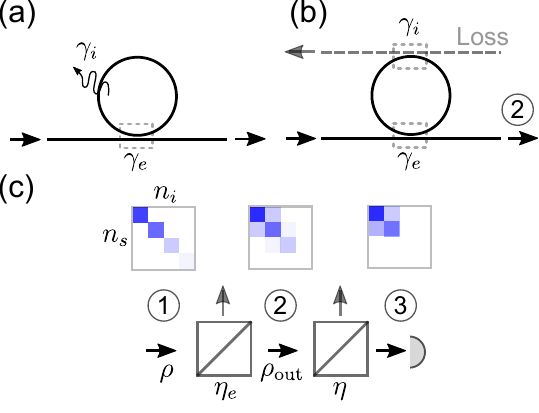}
    \caption{(a) Sketch of a lossy microresonator in which photons can decay in the waveguide at a rate $\gamma_e$ and being lost with a rate $\gamma_i$. (b) Equivalent model of the lossy resonator shown in panel (a), in which the losses are accounted by coupling the resonator with the auxiliary Loss waveguide. (c) Series of transformations that the TMS $\rho$ undergoes from generation to detection. (b). The three panels above the beamsplitters with transmittivity $\eta_e$ and $\eta$ sketch the joint probability distribution of the signal and the idler photon numbers $n_s$ and $n_i$ at the points (1), (2) and (3). 
    }
    \label{Fig_0}
\end{figure}
We aim to reconstruct the PND of two-mode squeezed light at the output of a lossy silicon nitride microresonator \cite{vaidya2020broadband}. The state is produced by pumping the resonator with a strong pulsed pump with central frequency $\omega_p$, which generates signal and idler conjugate beams centered at  frequencies $\omega_s$ and $\omega_i$, respectively, with $2\omega_p = \omega_s+\omega_i$. In order to understand, interpret and compare the experimental PND to that expected from theory, it is important to develop a model of SFWM in the resonator which takes into account both the pulsed nature of the pump, the presence of losses, and of the background noise. First, the presence of losses and spurious noise alters the photon number correlation between the squeezed modes. 
Second, we can relate Eq.(\ref{eq:pi_off}) to the Mandel operator describing a missed photodetection only in the case where the signal and idler photons are confined to a single and well defined spectral-temporal mode \cite{walls2012quantum}. The model should then be used to identify these modes, and validate to which extent SPM and XPM may alter their total number and distribution \cite{christ2011probing}.\\
We start to build our model from the geometry shown in Fig.\ref{Fig_0}(a). The resonator is coupled to a bus waveguide with a rate $\gamma_e$, and at the same time leaks photons into the external environment with a rate $\gamma_i$. 
This system is equivalent to a lossless resonator in which an additional waveguide, called  \emph{phantom channel}, is coupled to the cavity with a decay rate $\gamma_i$ \cite{vernon2015spontaneous}, which is schematically shown in Fig.\ref{Fig_0}(b). One way to solve the spontaneous problem is to write the hamiltonian as \mbox{$H=H_{\textrm{wg}}+H_{\textrm{ph}}+H_{\textrm{r}}+H_{\textrm{c}}+H_{\textrm{NL}}$}, where the first three terms describe the free-evolution of the waveguide(phantom) and resonator modes, $H_c$ mediates the coupling between the ring and the two channels and $H_{\textrm{NL}}$ is responsible for SFWM, self and cross-phase modulation (SPM/XPM) \cite{vernon2019scalable}. Alternatively, the linear part of $H$ can be diagonalized  to find the eigenmodes of the coupled ring-waveguide(phantom) system, and $H_{\textrm{NL}}$ can be written on this bases. These modes are called asymptotic fields and they span a continuum of frequencies \cite{liscidini2012asymptotic}. On these basis, the hamiltonian $H_{\textup{NL}}$ in the interaction picture can be written as \cite{quesada2022beyond}
\begin{equation}
\begin{gathered}
H_{\textup{NL}}(t) = \hbar \sum_{m,n}\left\{\int\Delta_s^{(m,n)}(\omega_s,\omega_s',t)a_{m}^{\dagger}(\omega_s)a_{n}(\omega_s')d\omega_s d\omega_s' +  \right.\\
+ \int\Delta_i^{(m,n)}(\omega_i,\omega_i',t)a_{m}^{\dagger}(\omega_i)a_{n}(\omega_i')d\omega_i d\omega_i' + \\
+\left. \left(\int \xi^{(m,n)}(\omega_s,\omega_i,t)a_{m}^{\dagger}(\omega_s)a_{n}^{\dagger}(\omega_i)d\omega_s d\omega_i + \textrm{H.c.}\right) \right\},
\end{gathered}
\label{eq:full_H}
\end{equation}
where the terms $\Delta^{(m,n)}_{s(i)}$ describe the frequency shifts induced by XPM (diagonal terms) and possible mode couplings (off-diagonal terms), while $\xi^{(m,n)}$ accounts for SFWM and depends on the strength of the nonlinearity, the internal pump power, the field enhancements at the pump, signal and idler wavelengths and on the mode volume of the resonator \cite{quesada2022beyond}. The indices $(m,n)=\{\textrm{wg},\textrm{ph}\}$ indicate the channel in which photons are scattered in the asymptotic limit of $t\rightarrow \infty$, when all nonlinear interactions have already occurred. In writing Eq.(\ref{eq:full_H}), the pump is treated as a classical field and pump-depletion is neglected. It is shown in Appendix D that the following relations hold
\begin{equation}
    \frac{\xi^{\textrm{wg,ph}}}{\xi^{\textrm{wg,wg}}} = \frac{\xi^{\textrm{ph,wg}}}{\xi^{\textrm{wg,wg}}} = \sqrt{\frac{\gamma_i}{\gamma_e}}, \quad \frac{\xi^{\textrm{ph,ph}}}{\xi^{\textrm{wg,wg}}} = \frac{\gamma_i}{\gamma_e}, \label{eq:xi_ratio}
\end{equation}
and identical properties relate the quantities $\Delta^{(m,n)}_{s(i)}$. In Eq.(\ref{eq:xi_ratio}) we assumed that the decay rates $\gamma_e$ and $\gamma_i$ are the same at the pump, signal and idler frequencies. One can show (see Appendix D) from Eq.(\ref{eq:xi_ratio}) and Eq.(\ref{eq:full_H}) that $H$ describes a TMS impinging on a beamsplitter with transmittivity $\frac{\gamma_e}{\gamma_e+\gamma_i}=\eta_e$. This equivalent picture is shown in Fig.\ref{Fig_0}(c).  
We can embed the loss $\eta_e$ on the transmitted photons in the detection efficiency $\eta'=\eta_e\eta$, and focus on the PND of the state immediately at the output of the lossless resonator (point (1) in Fig.\ref{Fig_0}(c)).  The quadratic hamiltonian in Eq.(\ref{eq:full_H}) generates a multimode squeezed vacuum state over a set of $\mathcal{K}$ orthogonal modes $A^{(s)}_k$ and $A^{(i)}_k$, called the Schmidt modes  \cite{christ2011probing,quesada2022beyond}. We can then write the state $\ket{\Psi}$ at the output of the resonator (point (1) in Fig.\ref{Fig_0}(c)) as
$\ket{\Psi} = \otimes_{k=1}^{\mathcal{K}} S_k(r_k)\ket{\textrm{vac}}$,
where $S_k=\exp\left(r_k {A_k^{(s)}}^\dagger {A_k^{(i)}}^\dagger -\textrm{H.c.} \right)$ is the two-mode squeezing operator associated to the $k^{\textup{th}}$ Schmidt mode and $r_k$ is the corresponding squeezing parameter. If the pump, the signal and the idler quality factors $Q=\frac{\omega}{(\gamma_e+\gamma_i)}$ are equal, and if the pump pulse duration is shorter than the dwelling time of photons within the cavity, the Schmidt number $K$ is $K\sim1.075$ \cite{vernon2017truly}. These conditions are satisfied in our experiment, so we assume to work in the approximation of one Schmidt mode. Under this hypothesis, the PND is given by  $\rho^{\textrm{SFWM}}_{n n'} = \left (\frac{\tanh^{n}(r)}{\cosh(r)} \right )$ \cite{quesada2022beyond}.
It follows that the marginal PND of the signal and the idler beam is thermal with a mean photon number of \mbox{$\langle n_s \rangle = \langle n_i \rangle =\langle n \rangle  = \sinh^2(r)$}. \textcolor{black}{At low powers, the relation between the squeezing parameter and the input power $P$ is linear, i.e., $r=aP$. Furthermore, when $aP\ll1$, we have that $\langle n \rangle \sim aP^2$, which implies that the rate of signal and idler photons generated by SFWM quadratically grows with the pump power. This breaks at higher power, where the spectral distribution of the signal and the idler photons modifies due to SPM/XPM shifts and to time ordering effects in the state evolution governed by the hamiltonian in Eq.(\ref{eq:full_H}) \cite{triginer2020understanding,quesada2022beyond,vendromin2024highly}. In this strongly nonlinear regime, no simple relation holds between $r$ and $P$.} 
The real PND should take into account 
the contribution of spurious noise photons that can be generated in the resonator due to spontaneous Raman scattering or fluorescence, whose rate scales as $\sim b_{s(i)}P$, in which $b_{s(i)}$ describes the efficiency of the parasitic process with the pump power. This contribution is uncorrelated from that of SFWM, and modifies the density matrix $\rho=\ket{\Psi}\bra{\Psi}$ as \mbox{$\rho\rightarrow \rho\otimes\rho^{\textrm{th,s}}\otimes\rho^\textrm{th,i}$}, where $\rho^{\textrm{th,s(i)}}$ is the density matrix of a thermal state with mean photon number $\langle n_{\textrm{s(i),th}}\rangle=b_{s(i)}P$. 
The modified PND $\rho_{nn'}$ is the convolution of the three uncorrelated PNDs
\begin{equation}
    \rho_{nn'} = \sum_{k,q,p} \rho^{\textrm{SFWM}}_{kk}\rho^{\textrm{th,s}}_p\rho^{\textrm{th,i}}_q\delta_{n,k+p}\delta_{n',k+q}. \label{eq:convolution}
\end{equation}
The PND at the output of the lossy resonator (point (2) in Fig.\ref{Fig_0}(b,c))  $\rho_{nn'}^{\textrm{out}}$ is given by \mbox{$\textrm{Tr}\big[U_{\textrm{BS}}(\rho\otimes\rho^{\textrm{th,s}}\otimes\rho^\textrm{th,i})U_{\textrm{BS}}^\dagger\ket{nn'}\bra{nn'}\big]$}, where $U_{\textrm{BS}}$ is the generator of the beamsplitter transformation with transmittivity $\eta_e$. With reference to Fig.\ref{Fig_0}(b,c) and by using the method described in Section \ref{sec:theory}, $\rho_{nn'}^{\textrm{out}}$ can be reconstructed by setting the losses to $\eta$. Indeed, these are the losses experienced by the signal and the idler beam from point (2) in Fig.\ref{Fig_0}(c,b) to the detectors.  On the other hand, $\rho_{nn'}$ is retrieved by setting the losses to $\eta\eta_e$ (point (1) in Fig.\ref{Fig_0}(c)). The expression in Eq.(\ref{eq:convolution}) strictly holds only for $\rho_{nn'}$, because it implies a perfect photon number correlation between the signal and the idler in the limit of vanishing noise.
\section{\label{sec:device} Device and experimental set-up}
\begin{figure}[t!]
    \includegraphics[scale = 0.83]{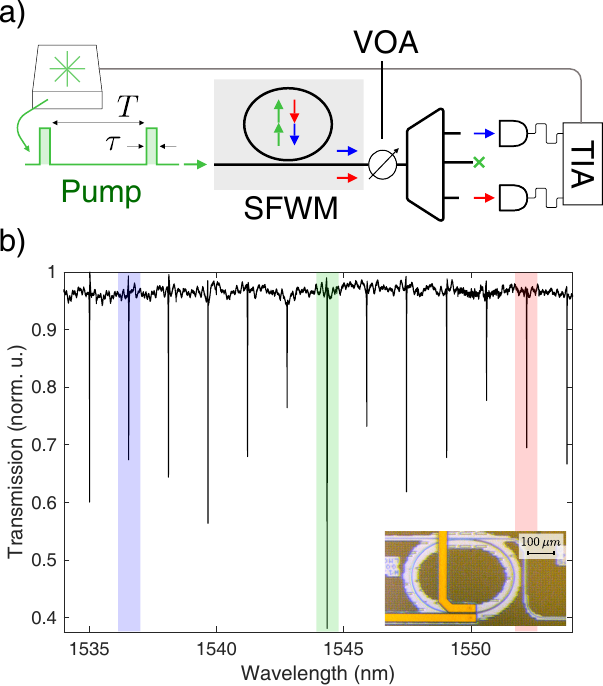}
    \caption{(a) Sketch of the experimental setup. The pump (green) pulse  duration is $\tau$ and the repetition rate $1/T$. The signal and idler beams are shown with red and blue arrows. VOA: variable optical attenuator, TIA: time-interval analyzer. (b) Transmission spectra of the microresonator. The pump, signal and idler resonances involved in the SFWM process are highlighted in green, red and blue respectively. The bottom right inset shows an optical microscope image of the device.}
    \label{Fig_1}
\end{figure}
Our device is a silicon nitride microresonator of cross section that is coupled to a single bus waveguide. Details of sample fabrication and device geometry are provided in Appendix A. We use a tunable laser to probe the transmission spectrum of the resonator, which is reported in Fig.\ref{Fig_1}(b). The free spectral range (FSR) is about  $200$ GHz. Two-mode squeezed light is generated at the signal and idler resonance wavelengths $\lambda_s = 1552.3$ nm and $\lambda_i = 1536.7$ nm by resonantly pumping the device at a wavelength of $\lambda_p=1544.5$ nm. The loaded Q of these resonances are $Q_s\sim Q_i = 2.6\times 10^5$ and  $Q_p = 3.3\times 10^5$. From the extinction of the resonances, we estimated an intrinsic quality factor of $Q_{\textrm{int}}\sim 3.5\times10^6$, and an escape efficiency \mbox{$\eta_e =(Q_{\textrm{int}}-Q)/Q_{\textrm{int}}\sim0.926$}.
The generation of squeezed light from the microresonator is investigated using the setup sketched in Fig.\ref{Fig_1}(a).
A continuous wave pump laser is passed through an electro-optic amplitude modulator(EOM) which carves pulses of duration $\tau = 300$ ps at a repetition rate of $2.5$ MHz. After the EOM, the pump is amplified by an Erbium-Doped Fiber Amplifier (EDFA). We stabilize the pump power with high precision ($<0.01\%$ rms of relative power fluctuation) by using a variable optical attenuator and a closed-loop optical feedback controlled by a field-programmable gate array (FPGA). A fiber  polarization controller is used to set the polarization to TE, which maximizes the SFWM efficiency. We suppress the laser background noise at the signal and idler wavelengths with a passband filter. Light is coupled in and out from the chip using two UHNA4 fibers and an index matching gel for improved coupling efficiency and mechanical stability \textcolor{black}{($1.0(5)$ dB loss/facet)}. The output is passed through a VOA which controls the overall loss $\eta$ of the system. The VOA attenuation $\eta_{\textrm{VOA}}$ (expressed net of the insertion loss) is calibrated with an accuracy below $2\%$ (including the polarization dependent loss) in the spectral range $[1500-1600]$ nm. The signal and the idler modes are separated by two Dense Wavelength Division Multiplexing (DWDM) modules with a bandwidth of $100$ GHz. Each filter is followed by a tunable fiber Bragg grating with a passband of $12.5$ GHz to reduce the in-band noise. 
Photons are detected by two superconducting nanowire single-photon detectors (SNSPD, \textcolor{black}{from PhotonSpot}). 
At $\eta_{\textrm{VOA}}=1$, the losses from the chip to the detector are $3.5(5)$ dB for the signal and the idler channels, including the fiber-to-chip coupling losses, the detector efficiency of $85\%$ and the escape efficiency $\eta_e$.
\section{\label{sec:model_validation}Model validation and limits of application}
\begin{figure}[t!]
    \includegraphics[scale = 0.43]{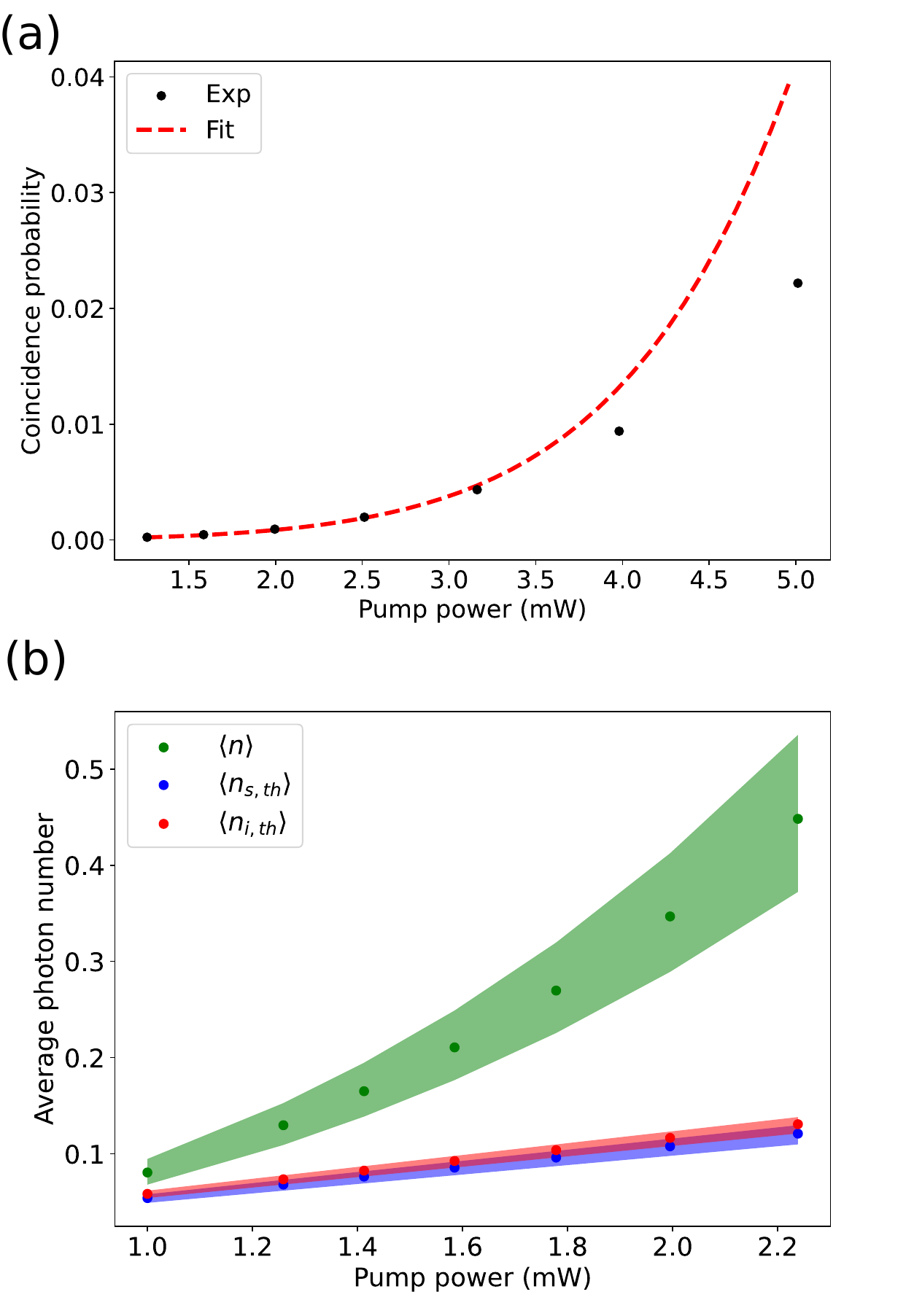}
    \caption{(a) Coincidence click probability (black dots) between the signal and the idler detector as a function of the input pump input power. The red dashed line is calculated from model in Eq.(\ref{eq:convolution}) and fits the data for $P<2.5$ mW. (b) Average number of photons per pulse from squeezing ($\langle n \rangle$) and from the thermal background in the signal ($\langle n_{\textrm{s,th}} \rangle$) and the idler mode  ($\langle n_{\textrm{i,th}} \rangle$). Confidence bounds covering one standard deviation are shown with colored regions. These are calculated from the uncertainty $\Delta\eta$ in the measurement of the losses, which is $\Delta\eta = 0.5$ dB.}
    \label{Fig_2}
\end{figure}
The model of the PND in Eq.(\ref{eq:convolution}) holds as long as the intra-cavity mean photon number $\langle n_{s(i)} \rangle$ is much lower than the number of pump photons, i.e., in the undepleted pump approximation. In our work, this approximation is well justified since, as it will be shown later, $\langle n_{s(i)} \rangle\sim 1$. However, the linear relation between the squeezing parameter $r$ and the input pump power $P$ holds only when SPM and XPM effects are negligible. This is validated by measuring the joint click probabilities $p_{11}$ as a function of the average power coupled to the chip. 
For each power, we maximize the count rate by  adjusting the wavelength of the pump laser to compensate the SPM and XPM induced shifts on the cold resonance frequencies. In Fig.\ref{Fig_2}(a) we show the measured coincidence probability $p_{11}$ as a function of the pump power $P$. 
We fit $p_{11}$ using Eq.(\ref{possibili_4_eventi}) with the PND given by Eq.(\ref{eq:convolution}), leaving $a$ and $b_{s(i)}$ as free parameters. We found that the $R^2$ of the fit decreases when click-probabilities at input powers higher than $2.5$ mW are included in the fit. We then limit the fit to the range $1$ mW $\le P \le2.5$ mW, and show the result in Fig.\ref{Fig_2}(a).
We see that the model correctly predicts the experimental data for input powers \mbox{$<2.5$ mW}, while at higher powers it overestimates the generation rate. It is clear that for \mbox{$P>3$ mW}, the linear relationship between the squeezing parameter and the input power no longer holds. The behavior is different from that of a CW pump, in which the SPM and XPM induced resonance shifts can be fully compensated by adjusting the pump frequency, and the generation rate scales with the square of the input power \cite{vernon2015strongly}. 
Using the parameters $a$ and $b_{s(i)}$, we calculated the mean number of photons per pulse from squeezing ($\langle n \rangle = \sinh^2(aP)$) and noise ($\langle n_{\textrm{s(i),th}} \rangle = b_{s(i)}P$), which are shown in Fig.\ref{Fig_2}(b). 

\section{Reconstruction of the photon number distribution}
\begin{figure*}[t!]
    \centering\includegraphics[scale = 0.28]{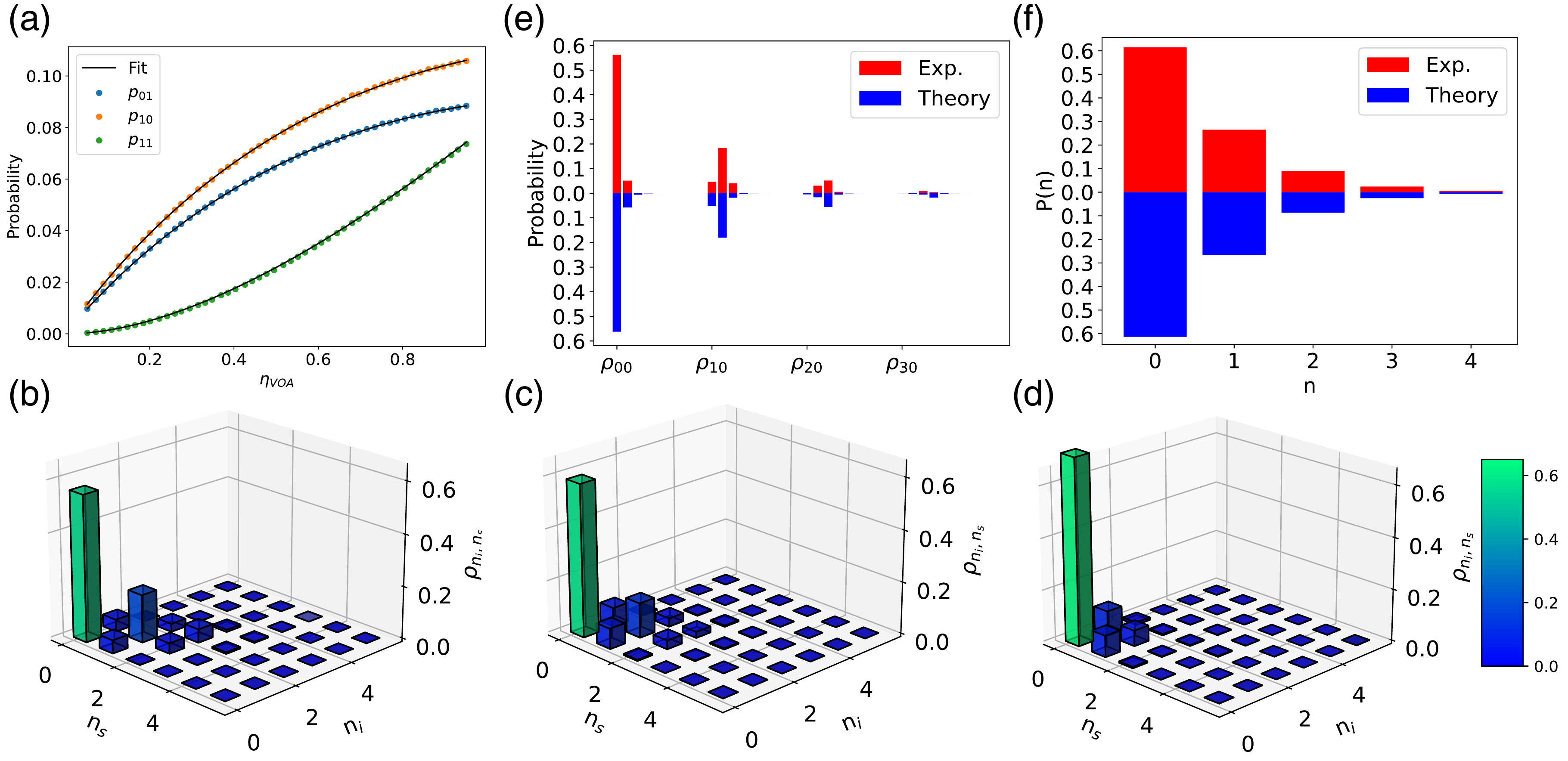}
    \caption{(a) Single ($p_{01}$, $p_{10}$) and joint ($p_{11}$) click probabilities (dots) for different VOA attenuations $\eta_{\textrm{VOA}}$  at a pump power of $2.2$ mW. The black line is a fit of the data using the model in Eq.(\ref{eq:convolution}) and Eq.(\ref{possibili_4_eventi}). (b) Reconstructed photon number distribution $\rho_{\textrm{EM}}$ at the output of the resonator for $2.2$ mW of input pump power, \textcolor{black}{corresponding to an estimated on chip squeezing level of $6.2(2)$ dB.} (c,d) Same as in (b), but reconstructed at the output of the chip (c) and at the detectors \textcolor{black}{(squeezing level of $2.2(2)$ dB)} (d). (e) Comparison between the experimental PND shown in panel (b) and the simulation with highest fidelity. (f) Comparison between the experimental marginal photon number distribution of the signal and 
    the simulated marginal distribution with highest fidelity.}
    \label{Fig_3}
\end{figure*}
The PND is reconstructed using the method presented in Sec.\ref{sec:theory}. We implemented $M=50$ measurement settings by varying the VOA transmission $\eta_{\textrm{VOA}}$ from a minimum of $0.05$ to a maximum of $0.95$ in equal steps. The total transmission is then $\eta=\eta_{\textrm{exp}}\eta_{\textrm{VOA}}$, where  $\eta_{\textrm{exp}}$ is transmission $\eta$ measured with the VOA set at the minimum of attenuation. For each setting $\mu$, we extracted the four click-probabilities $p_{xy}(\eta_{\mu})$ by collecting  $12.5$ millions of events. The  probabilities are input to the EM algorithm and a uniform PND is used as an initial uninformative ansatz. The algorithm is stopped when the variation in relative error $\frac{\Delta\epsilon}{\epsilon}$ in Eq.(\ref{error_formula}) is below $10^{-3}$. 
The dimension of the Hilbert space is truncated to a dimension of $10$.
In Fig.\ref{Fig_3}(a) we show the measured click probabilities $p_{xy}$ as a function of the VOA transmission for an input power of $2.5$ mW. We also show the predicted click probabilities $p_{xy}(\eta)\{\rho_{\textrm{EM}}\}$ given the PND $\rho_{\textrm{EM}}$ at the last iteration step of the EM. These are in a very good agreement with the experiment. The reconstructed PND is reported in Fig.\ref{Fig_3}(b) and shows strong photon number correlations between the signal and the idler mode, as expected for a TMS. The off-diagonal elements in the PND matrix $\rho_{\textrm{EM}}$ arise from two contributions. The first is the presence of uncorrelated noise photons in the signal and idler arms, mainly due to spontaneous Raman scattering. This represents the dominant contribution. The second could be due to an imperfect estimation of the losses $\eta_{\textrm{exp}}$. If the latter are higher than those used to retrieve $\rho_{\textrm{EM}}$, i.e.  $\eta_{\textrm{exp}}'<\eta_{\textrm{exp}}$, then the reconstructed PND will describe the statistics of the state generated by the (lossless) resonator and subjected to an extra-attenuation of $\frac{\eta_{\textrm{exp}}'}{\eta_{\textrm{exp}}}$. This loss spoils the perfect photon number correlation in absence of noise. \\
With our method, we can easily track the changes of the PND from the chip to the detectors. In Fig.\ref{Fig_3}(c,d) we report respectively the reconstructed PND immediately out of the chip and at the detectors. The two PNDs are obtained by substituting $\eta_{\textrm{exp}}\rightarrow \frac{\eta_{\textrm{exp}}}{\eta_{\textrm{chip}}}$ and $\eta_{\textrm{exp}}\rightarrow 1$ in the EM algorithm, where $\eta_{\textrm{chip}} = 0.8$ is the chip-to-fiber coupling efficiency. Compared to the on-chip PND, the photon number correlation is gradually lost by adding losses, and tends to that of the product of two uncorrelated thermal states. This is consistent with the fact that for a TMS the variance $V_{\Delta n}$ of the photon number difference $\Delta_n = n_s-n_i$ decreases with the losses as $V_{\Delta n}=\eta\langle n_s+n_i \rangle$ \cite{vaidya2020broadband}.  
Since the EM reconstructs $\rho_{\textrm{EM}}$ without relying on any model but solely from the raw click probabilities, it is important to calculate the fidelity $\mathcal{F}$ between $\rho_{\textrm{EM}}$ and the physics-driven model distribution in Eq.(\ref{eq:convolution}). We obtained a maximum fidelity of $\mathcal{F}=0.98$ when $r=0.63$, $\langle n_s \rangle = 0.11$ and $\langle n_i \rangle = 0.10$. The comparison between $\rho_{\textrm{EM}}$ and the theoretical distribution is shown in Fig.\ref{Fig_3}(e). The remarkable agreement between the two indicates that $\rho_{\textrm{EM}}$ is a reliable reconstruction of the PND of the state, and that we can use Eq.(\ref{eq:convolution}) to infer the squeezing parameter $r$. We also apply the EM to reconstruct the marginal photon number distribution of the signal(idler) arm, and evaluated the fidelity with the distribution obtained by marginalizing Eq.(\ref{eq:convolution}) with respect to the idler(signal) photon number. The comparison between the two PND is reported in Fig.\ref{Fig_3}(f) and indicates a very good agreement, with a fidelity of $\mathcal{F}=0.999$. 
As a side remark, we noticed the emergence of artifacts in the reconstructed PND for input powers higher than $3.25$ mW. We traced this fact back to an overfitting problem, which becomes relevant at high powers due to the increased number of not-vanishing elements in the PND and to the high sensitivity of those elements to experimental noise.

\subsection{Calculation of the mean photon number, squeezing parameter and NRF from the PND}
\begin{figure}[t!]
    \includegraphics[scale = 0.45]{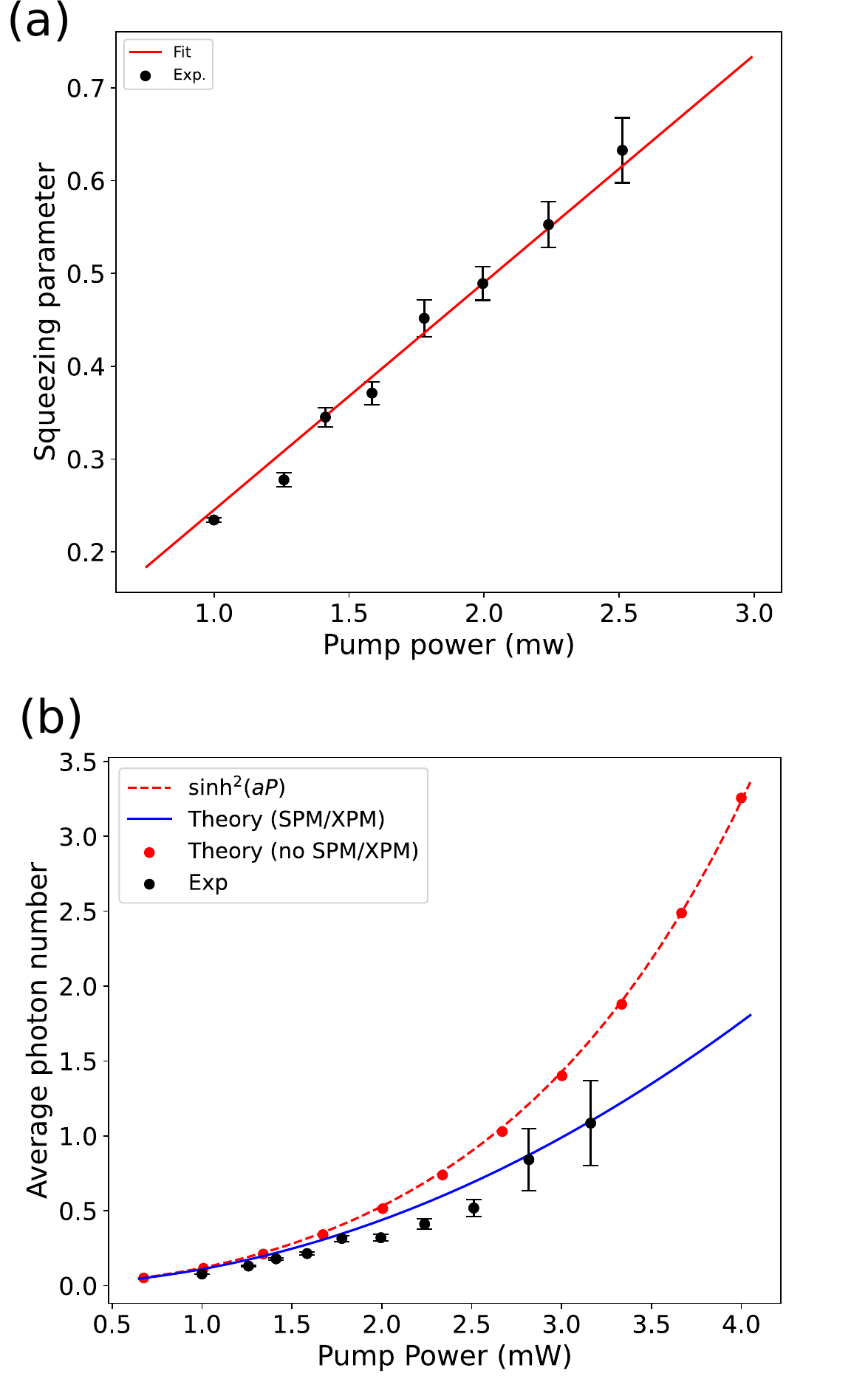}
    \caption{(a) Squeezing parameter $r$ extracted from the PND as a function of the input pump power (black dots). The red line is a linear fit of the data. (b) Simulated average number of photons per pulse $\langle n \rangle$ for different input powers. The blue dots consider the effects of SPM and XPM, while the red dots are obtained with SPM and XPM switched off. The red dashed line is a fit of the data without SPM and XPM using the function $\sinh^2(aP)$. The black dots are the values of $\langle n \rangle$ extracted from the PND in the experiment.} 
    \label{Fig_4}
\end{figure}

The measurement of the PND allows us to estimate important metrics. The first one we calculate is the squeezing parameter $r$. We reconstructed the PND for different input pump powers, and for each we find the values of $r$ and $\langle n_{\textrm{s(i),th}}\rangle$ that maximize the fidelity with the distribution in Eq.(\ref{eq:convolution}). These are shown in Fig.\ref{Fig_4}(a) for $P<2.5$ mW, for which Eq.(\ref{eq:convolution}) is a faithful model. 
The result shows that $r$ scales linearly with $P$, which agrees with the theory and is consistent with the power scaling of the click probabilities shown in Fig.\ref{Fig_2}(a), with the important difference that in the former case no prior relations between the two variable is assumed. We can exploit the PND to calculate the mean photon number $\langle n \rangle$ per pulse. This quantity had already been calculated with the model in Eq.(\ref{eq:convolution}) (see Fig.\ref{Fig_2}(b)), but the results were justified only in the low power regime. In Fig.\ref{Fig_4}(b) we show $\langle n \rangle$ as a function of the input power up to $P=3.25$ mW, i.e., the highest power for which no artifacts were observed in the reconstructed PNDs. To validate the reliability of the reconstructed PND and of the value of $\langle n \rangle$ in this extended power range, we numerically solved the stochastic master equation for the signal and idler modes under the action of the hamiltonian in Eq.(\ref{eq:full_H}) and subjected to continuous photodetection of both the signal and idler modes \cite{wiseman2009quantum}. The simulation details are reported in Appendix E. From the simulated PND, we calculated the mean number of signal(idler) photons scattered into the bus waveguide for different input powers $P$, which we shown in Fig.\ref{Fig_4}(b). 
The experimental values of $\langle n \rangle$ are in very good agreement with the simulation. The full list of the fidelities between the simulated and the experimental PND is reported in Table \ref{Table_2} of Appendix E. We also calculated $\langle n \rangle$ by switching off SPM and XPM, and overlaied the results in Fig.\ref{Fig_4}(b). The effect of SPM and XPM is to decrease the average number of photons generated by SFWM, which explains why the relation $r=aP$ overestimates the coincidence click probabilities shown in Fig.\ref{Fig_2}(a). 
\begin{figure}[t!]
    \includegraphics[scale = 0.4]{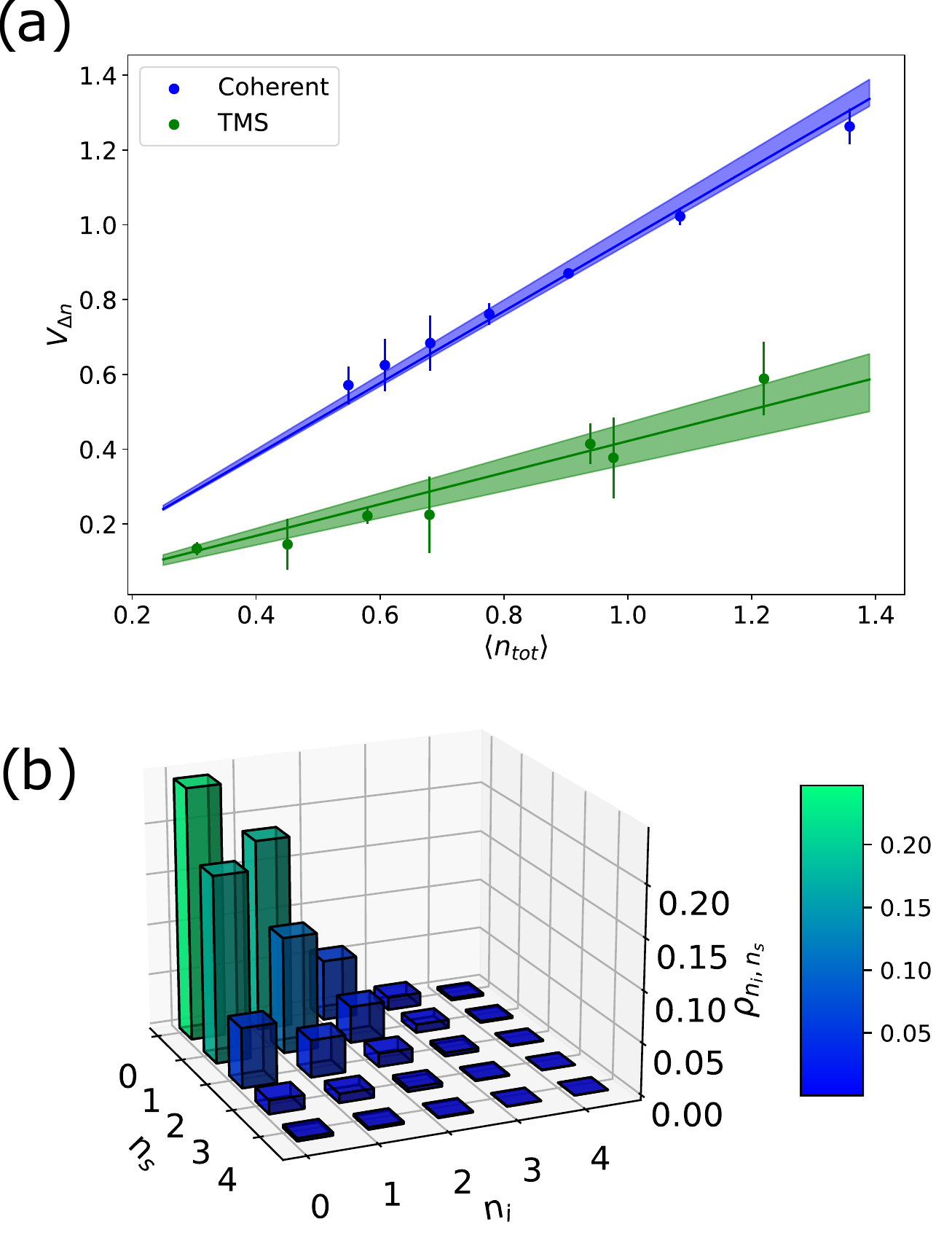}
    \caption{(a) Variance $V_{\Delta n}$ of the photon number difference between the signal and the idler arm as a function of the average photon number $\langle n_{\textrm{tot}} \rangle$. The blue dots refer to a bipartite coherent state while the black dots to the squeezed light from the resonator. Solid lines are linear fit of the data. Confidence bounds covering one standard deviation are shown with colored regions. These are calculated from the uncertainty $\Delta\eta$ in the measurement of the losses, which is $\Delta\eta = 0.5$ dB. (b) Reconstructed PND for a bipartite coherent state with an average photon number of $\langle n \rangle=1$. } 
    \label{Fig_5}
\end{figure}
We noticed that, similarly to the CW case, there is an optimal and power dependent pump frequency detuning from resonance for which $\langle n \rangle$ is maximized. Intuitively, the nonlinear resonance shifts can not be totally compensated because the pump detuning is kept fixed, while the SPM and XPM-induced resonance shifts follow the temporal variations of the pump intensity in the resonator. As shown in Fig.\ref{Fig_4}(b), the mean number of generated photons per pulse scales as $\langle n \rangle = \sinh^2(aP)$ if SPM and XPM effects are not cosidered. We used the numerical simulation to calculate the time-integrated unheralded $\bar{g}_2$ (see Appendix E) of the signal/idler modes, which is connected to the Schmidt number $K$ as $K=g_2-1$ \cite{christ2011probing}. We found that up to $4$ mW of input power, its value changes by less than $1\%$ from that at low power ($\bar{g}_2=1.89$), implying that SPM and XPM impact the squeezing strength more than the number of Schmidt modes. \\
As a last step, from the PND we calculated the variance $V_{\Delta n}$ of difference between the signal and the idler photon numbers, which is given by $V_{\Delta n} = V_{n_s} + V_{n_i} - 2 \big( \langle n_s n_i \rangle -  \langle n_s \rangle  \langle n_i \rangle \big)
$, where $V_{n_{s(i)}}$ is the variance of the signal(idler) photon number. For an ideal TMS subjected to a loss $\eta$, this quantity is equal to $V_{\Delta n}=\eta(\langle n_s\rangle +\langle n_i) \rangle = \eta \langle n_{\textrm{tot}} \rangle$, while for a coherent and a thermal state is respectively equal to $V_{\Delta n} = \langle n_{\textrm{tot}}\rangle$ and $V_{\Delta n} = \langle n_{\textrm{tot}} \rangle + \langle n_s \rangle^2 + \langle n_i \rangle ^2$ \cite{loudon2000quantum,vaidya2020broadband}. A reduction of the slope between $V_{\Delta n}$ and $\langle n_{\textrm{tot}} \rangle$ from unity indicates correlations between the number of signal and idler photons. The ratio $\frac{V_{\Delta n}}{\langle n_{\textrm{tot}}\rangle}$ is the NRF, or photon number difference squeezing. A value below unity is considered a metric of non-classicality of the state \cite{aytur1990pulsed}. The variance $V_{\Delta n}$ as a function of $\langle n_{\textrm{tot}} \rangle$ is calculated from the PND reconstructed immediately at the output of the lossy resonator, and is reported in Fig.\ref{Fig_5}(a). The slope of the linear fit is $0.42(2)$, corresponding to a photon number difference squeezing of $-3.8(2)$ dB. The expected value of the NRF calculated from the escape efficiency of the resonator is $10\log(1-\eta_e)\sim-11$ dB. However, the  correlation between the signal/idler photon number difference is reduced by the background noise. Indeed, by calculating $\frac{V_{\Delta n}}{\langle n_{\textrm{tot}}\rangle}$ using the the PND in Eq.(\ref{eq:convolution}) we obtain $\frac{V_{\Delta n}}{\langle n_{\textrm{tot}}\rangle}=0.4(1)$ ($-4.0(1)$ dB of photon number difference squeezing), which is consistent with the NRF value in Fig.\ref{Fig_5}(a). We complement the analysis by comparing the NRF of the TMS from that of a coherent state. The latter is obtained by passing the attenuated pulsed pump laser into a 50/50 fiber beamsplitter and by reconstructing its PND at different powers. An example of such reconstruction is shown in Fig. \ref{Fig_5}(b). The variance $V_{\Delta n}$ as a function of $\langle n_{\textrm{tot}} \rangle$ is shown in Fig.\ref{Fig_5}(a). The calculated NRF is $0.96(2)$ and lies well above that of the TMS. 

\section{\label{sec:concl}Conclusions} 
In this work we reconstructed the PND of pulsed squeezed light generated from a silicon nitride microresonator without the use of photon number resolving detectors. The method exploits the nonlinear response of on-off photodetectors to photon number states undergoing variable attenuation. When SPM and XPM effects have negligible impact on the rate of photon pairs generated by SFWM, the reconstructed photon statistics is well described by the convolution of a TMS and two uncorrelated thermal fields. The latter account for spurious noise photons generated by the resonator, and mainly arise from spontaneous Raman scattering. 
At low power, we found that the squeezing parameter scales linearly with the pump power. At higher input powers, the squeezing strength is limited by SPM and XPM. In contrast to the CW case, it is not possible to fully compensate the nonlinear resonance shifts by adjusting the pump frequency. 
The PND are used to calculate the mean photon number per pulse, the squeezing parameter and the noise reduction factor. 
The PNDs are characterized up to $\sim 1.2$ photons/pulse, through which we extracted an on-chip squeezing level of $6.2(2)$ dB and a noise reduction factor of $-3.8(2)$ dB.
This approach gives reliable results when the number of parameters in the PND is not too large, which may depend on several factors. The most relevant are the complexity of the PND, the accuracy and reproducibility of the VOA settings, the stability of the measuring apparatus over time, the overall losses and the finite sampling size. In our case, we found that reliable results are obtained for a Hilbert space dimension $<6\times6$. We expect that larger dimensions can be reached by complementing this method with the spatial and temporal multiplexing strategies used in pseudo-PNR.
\subsection*{Acknowledgements}
D.B acknowledges the support of Italian MUR and the European Union - Next Generation EU through the PRIN project number F53D23000550006 - SIGNED. E.B., M.B., M.B, M.G and M.L. acknowledge the PNRR MUR project PE0000023-NQSTI. All the authors acknowledge the support of Xanadu Quantum Technology for providing the samples and the useful discussions.

\section*{Appendix A: Device geometry and fabrication}
\label{sec:appendix_A}
The device is fabricated on a silicon nitride photonic chip in a standard multi-project wafer run by Ligentech. All the waveguides have width of \mbox{1.75 $\mu$m} and a thickness of \mbox{800 nm}, and are covered by a silicon dioxide cladding of thickness $4.1\,\mu$m. The gap between the resonator and the bus-waveguide is \mbox{0.4 $\mu$m}, and the corresponding coupling coefficient is $\kappa^2=0.016$. The resonator is made by two U-shaped Euler bends of length $360\,\mu$m, which are designed to minimize the bending loss, thus maximizing the intrinsic quality factor of the devices. The linear propagation losses of the Euler have been found comparable to those of the straigth waveguide sections, which are $\sim0.11$ dB/cm. A metallic heater of Titanium Nitride of $0.4\,\mu$m thickness and $2\,\mu$m width is placed $1.7\,\mu$m above the waveguides. It allows to change the resonance wavelength through the thermo-optic effect.
\begin{figure}[h!]
   \centering \includegraphics[scale = 0.38]{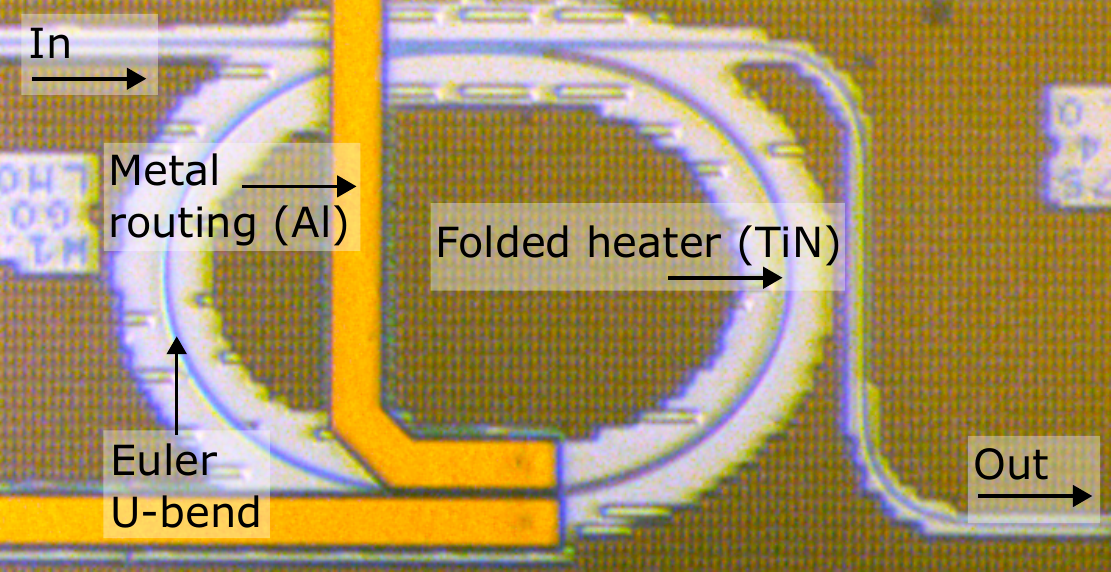}
    \caption{Optical microscope image of the silicon nitride microresonator used in this work. Indicated are the folded TiN heater used to locally heat the waveguide underneath, the Aluminium routing layer from the TiN heaters to the metal pads, the input and the output optical ports.} 
    \label{Fig_AP}
\end{figure}
\section*{\label{sec:appendix_B} Appendix B: reconstruction of single mode coherent and thermal states}
\begin{figure}[b!]
    \includegraphics[scale = 0.85]{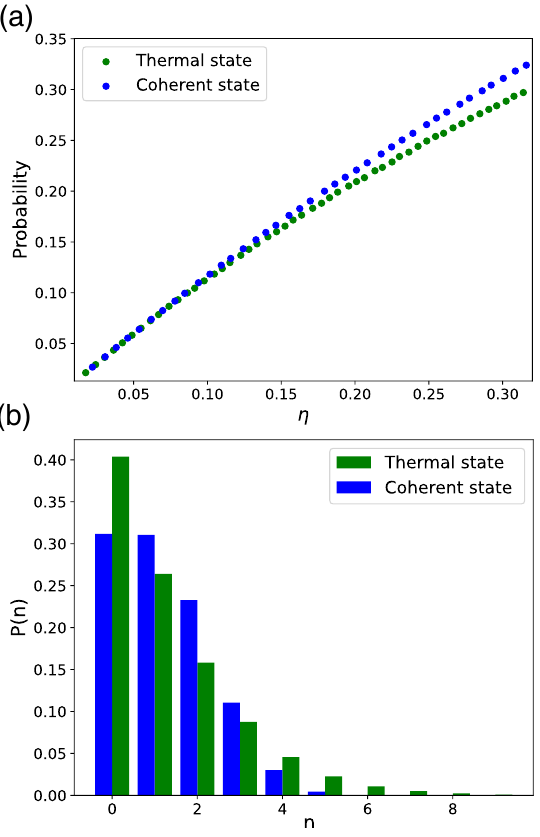}
    \caption{(a) Measured off-detection probabilities as a function of the system loss $\eta$ for a thermal and a coherent state with the same average photon number of $\langle n \rangle \sim 1.2$. (b) Reconstructed PND of the two states.} 
    \label{Fig_S1}
\end{figure}

We first validated the method presented in Sec.2 by reconstructing some known optical states, which we chose to be a single mode coherent state and a thermal state. To highlight that the measurement is sensitive to the full PND and not just to its first order moment, i.e. the average photon number $\langle n \rangle$, we aimed to reconstruct states with the same $\langle n \rangle\sim 1.2$. The coherent state is obtained by attenuating the pump laser, while the second from the unheralded signal beam generated by SFWM in the microresonator. Figure \ref{Fig_S1}(a) reports the probability of missed detections $p_0(\eta)$ of the two states as a function of the variable detection efficiency $\eta$. 
We notice that up to $\eta\sim 0.15$, the click probabilities are almost identical, which agrees with the fact that for $\eta_{\textrm{max}}\ll 1$ the probability of an missed detection is $p_0=\eta \langle n \rangle$. As $\eta$ increases, the photodetection statistics of the two states become distinguishable. This can be appreciated also from the reconstructed PND of the two states, which are shown in Fig.\ref{Fig_S1}(b). The experimental PNDs are then compared to their theoretical distributions, obtaining a fidelity of $\mathcal{F} = 0.997$ for the thermal state, and $\mathcal{F} = 0.998$ for the coherent state.
\newpage
\section*{\label{sec:appendix_C} Appendix C: The bipartite case}
The method presented in Sec.2 generalizes to the bipartite case, in which a single detector is used to sense each mode. The four possible photodetection events $p_{xy}$ occur with probabilities
\begin{equation} \label{possibili_4_eventi}
    \begin{split}
     &p_{00} (\eta) = \sum_{n,k} B_{\eta n} B_{\eta k} \rho_{nk},\\
     &p_{10} (\eta)  = \sum_{n,k}(1-  B_{\eta n}) B_{ \eta k} \rho_{nk},\\
     &p_{01} (\eta)  = \sum_{n,k}  B_{\eta n} (1- B_{ \eta k}) \rho_{nk},\\
     &p_{11} (\eta)  = 1- p_{00} (\eta) - p_{10} (\eta) - p_{01} (\eta),
    \end{split}
\end{equation}
in which the diagonal elements of the two-mode density matrix $\hat{\rho}$ are
\begin{equation}
    \rho_{nk} = \bra{nk}\hat{\rho}\ket{nk}, \quad \ket{nk} = \ket{n} \otimes  \ket{k}.
\end{equation}
The indices $(n,k)$ in Eq.(\ref{possibili_4_eventi}) range from $0$ to $N$, where $N$ is the dimension of the truncated Hilbert space.
We now define the vectorized PND $\boldsymbol{\rho}$ as 
\begin{equation} \label{formula_immagine_finale}
\boldsymbol{\rho} = (\rho_{00}, \rho_{01}, \dots,\rho_{0N}, \rho_{10},\rho_{11},\dots \rho_{N0},\rho_{N1},\dots \rho_{NN})^T,
\end{equation}
and the vector $\boldsymbol{p}$ collecting the $3M$ click-probabilities from the $M$ measurement settings as
\begin{equation} \label{matrice-vettore}
\boldsymbol{P} = (p_{00}^{\eta_1}, p_{00}^{\eta_2}, \dots , 
p_{00}^{\eta_M},
p_{01}^{\eta_1}, p_{01}^{\eta_2}, \dots ,
p_{01}^{\eta_M},
p_{10}^{\eta_1}, p_{10}^{\eta_2}, \dots
p_{10}^{\eta_M})^T.
\end{equation}
The system of equations in Eq.(\ref{possibili_4_eventi}) can be written in compact form as $\boldsymbol{p} = B\boldsymbol{\rho}$, in which
the $3M\times(N+1)^2$ matrix $B$ has elements
\begin{equation}
  B =
    \begin{cases}
     B_{\mu n} B_{\mu k} \qquad \quad\,\,\, \mu = 1, \dots, M\\
     B_{\mu n} (1- B_{\mu k}) \quad \mu = M +1, \dots, 2M\\
     (1- B_{\mu n}) B_{\mu k} \quad \mu = 2M +1, \dots, 3M
    \end{cases} 
\end{equation}
The EM step at the iteration $i$ becomes
\begin{equation}
    \rho_{p}^{(i+1)} = \rho_{p}^{(i)} \sum_{\mu = 1}^{3M} \frac{B_{\mu p} f_{\mu}}{ (\sum_{\nu} B_{\nu p}) p_{\mu} \{ \boldsymbol{\rho^{(i)}} \} },
    \label{eq:EM_bipartite}
\end{equation}
where we have defined the experimental frequencies as
\begin{equation}
    f_{\mu}=
    \begin{cases}
   f_{00} = n_{00 \mu} / n_{\mu} \quad \mu = 1, \dots, M\\
   f_{01} = n_{01 \mu} / n_{\mu} \quad \mu = M+1, \dots, 2M\\
   f_{10} = n_{10 \mu} / n_{\mu} \quad \mu = 2M+1, \dots, 3M\\
    \end{cases}
\end{equation}
where $n_{\mu}=R_\mu T_\mu$ is the total number of events, which is given by the repetition rate of the pump laser $R_\mu$ and the integration time $T_\mu$ of the $\mu$ measurement setting, while $n_{xy\nu}$ is the number of recorded events. In Eq.(\ref{eq:EM_bipartite}), we implemented the following mapping of the indices 
\begin{equation}
    \rho_{nk} \rightarrow, \rho_p \qquad p=1+k+n(1+N),
\end{equation}

\section*{\label{sec:appendix_D} Appendix D: Relation between the $\xi$ coefficients in the hamiltonian}
The function $\xi^{(m,n)}(\omega_s,\omega_i,t)$ in Eq.(7) of the main manuscript is equal to \cite{borghi2020phase}
\begin{equation}
    \xi^{(m,n)}(\omega_s,\omega_i,t) =  \mathcal{N}_p(\omega_s,\omega_i,t)\phi_{s,m} (\omega_s)\phi_{i,n} (\omega_i),
    \label{eq:xi_definition}
\end{equation}
in which
\begin{equation}
\begin{gathered}
\mathcal{N}_p(\omega_s,\omega_i,t) = \mathcal{N}'\int \phi_p(\omega_{p_1})\phi_p(\omega_{p_2})\\
\beta(\omega_{p_1})\beta(\omega_{p_2})e^{i\Delta \omega t}d\omega_{p_1}d\omega_{p_2},    
\end{gathered}
\label{eq:np_definition}
\end{equation}
and $\mathcal{N}'$ is a constant. In Eqs.(\ref{eq:xi_definition},\ref{eq:np_definition}),  $\phi_{p/s/i,m}$ are the field enhancements near the pump, signal and idler frequencies when the light is input to port \mbox{$m=\{\textrm{wg,ph}\}$}, $\beta$ is the spectral profile of the pump and \mbox{$\Delta\omega = \omega_s+\omega_i-\omega_{p_1}-\omega_{p_2}$}. Equation (\ref{eq:np_definition}) assumes that the relation between the incident pump field and the (slowly varying) internal field $E_p$ (normalized such that $[|E_p|^2]=\textrm{Watts}$) in the resonator is linear, i.e, $E_p\propto \beta\phi_p$. When SPM effects are not negligible, we have to replace $\beta\phi_p$ with the Fourier transform of the internal field $E_p(t)$. The latter satisfies the differential equation
\begin{equation}
    \frac{dE_p}{dt}=[-\gamma_{\textrm{tot,p}}+i(\Lambda|E_p|^2+\Delta_p)]E_p+i\sqrt{\frac{2\gamma_{ep}}{\tau_{\textrm{rt}}}}\beta(t), \label{eq:pump_eq}
\end{equation}
in which $\gamma_{\textrm{tot,p}}$ is the total decay rate of the field, $\gamma_{ep}$ is the coupling rate between the resonator and the bus waveguide, $\Delta_p=\omega_c-\omega_{p0}$ is the detuning of the central frequency $\omega_c$ of the pump laser from the cold resonance frequency of the pump $\omega_{p0}$ and $\tau_{\textrm{rt}}$ is the round-trip time of light in the resonator. The nonlinear coefficient $\Lambda$ accounts for SPM and is defined as
\begin{equation}
    \Lambda = \frac{\omega_p c n_2}{V_{\textrm{eff}}n_0^2}\tau_{\textrm{rt}}, \label{eq:Lambda}
\end{equation}
where $n_2$ is the nonlinear refractive index of silicon nitride, $n_0$ is the refractive index of the core material and $V_{\textrm{eff}}$ is the effective volume of the resonator. We can obtain $\phi_p$ in the low power regime by eliminating the SPM term in Eq.(\ref{eq:pump_eq}). This gives
\begin{equation}
    \phi_p(\omega) = \frac{i\sqrt{\frac{2\gamma_{ep}}{\tau_{\textrm{rt}}}}}{\gamma_{\textrm{tot}}-i(\Delta_p+\omega)}. \label{eq:fe_definition_p}
\end{equation}
Similarly, $\phi_{s(i),m(n)}$ are given by
\begin{equation}
\phi_{s(i),m(n)}(\omega) = \frac{i\sqrt{\frac{2\gamma_{s(i)}^{m(n)}}{\tau_{\textrm{rt}}}}}{\gamma_{\textrm{tot,s(i)}}-i(\omega_{s(i)}-\omega_{s(i)0})}, \label{eq:fe_definition_si}
\end{equation}
where $\gamma_{s(i)}^{m} = \gamma_{es(i)}$ if $m = \textrm{wg}$ and $\gamma_{s(i)}^{m} = \gamma_i$ if $m=\textrm{ph}$. From the definition in Eq.(\ref{eq:fe_definition_si}) we have that \mbox{$\phi_{\textrm{s(i),wg}} = \sqrt{\frac{\gamma_{es(i)}}{\gamma_i}}\phi_{\textrm{s(i),ph}}$}, which combined with Eq.(\ref{eq:xi_definition}) gives the relations in Eq.(8) in the main text, where $\gamma_{es} = \gamma_{ei} = \gamma_e $ has been used for simplicity. \\
The XPM induced by the pump to the signal and idler fields is described by the term 
\begin{equation}
\Delta_{s(i)}^{(m,n)} =   2\Lambda |E_p(t)|^2 \phi_{s(i),m}(\omega_{s(i)})\phi_{s(i),n}^{*}(\omega_{s(i)}')e^{i\delta_{s(i)}t},
\end{equation}
where $\delta_{s(i)} = \omega_{s(i)}-\omega_{s(i)}'$.
Armed with these tools, we can show that the hamiltonian in Eq.(7) of the main text is equivalent to 
that of a lossless resonator coupled to
a single bus waveguide with a rate $\gamma_e+\gamma_i$, in which the outgoing photons impinge on a beamsplitter with transmittivity $\eta_e = \frac{\gamma_e}{\gamma_e+\gamma_i}$. The nonlinear hamiltonian of the lossless resonator, in the interaction picture, is given by
\begin{equation}
\begin{gathered}
H_{\textup{NL}}(t) = \hbar \left\{\int\Delta_s^{\textup{(wg,wg)}}(\omega_s,\omega_s',t)a_{\textup{wg}}^{\dagger}(\omega_s)a_{\textup{wg}}(\omega_s')d\omega_s d\omega_s' +  \right.\\
+ \int\Delta_i^{\textup{(wg,wg)}}(\omega_i,\omega_i',t)a_{\textup{wg}}^{\dagger}(\omega_i)a_{\textup{wg}}(\omega_i')d\omega_i d\omega_i' + \\
+\left. \left(\int \xi^{\textup{(wg,wg)}}(\omega_s,\omega_i,t)a_{\textup{wg}}^{\dagger}(\omega_s)a_{\textup{wg}}^{\dagger}(\omega_i)d\omega_s d\omega_i + \textrm{H.c.}\right) \right\},
\end{gathered}
\label{eq:H_lossless}
\end{equation}
because once they are generated by SFWM, photons can only leave resonator from the bus waveguide. Those photons impinge on a beasmplitter with transmittivity $\eta_e$, which realize the transformation \mbox{$a_{\textup{wg}}\rightarrow \eta_e a_{\textup{wg}}+\sqrt{1-\eta_e}a_{\textup{ph}}$}, where $a^{\dagger}_{\textup{ph}}$ creates a photon in the phantom loss channel in Fig.1(b) of the main manuscript. By inserting this operator transformation into Eq.(\ref{eq:H_lossless}) and by using Eq.(8) we obtain the hamiltonian in Eq.(7) of the main manuscript, concluding our proof.
\section*{\label{sec:appendix_E} Appendix E: Stochastic master equation for the signal and idler fields}
The use of asymptotic fields to represent the Hamiltonian in Eq.(7) of the main text is useful  when SPM and XPM effects are negligible. In this case, the time dependence of the hamiltonian is solely contained in the frequency mismatch term in Eq.(\ref{eq:np_definition}), which is approximated by $\delta(\omega_s+\omega_i-\omega_{p_1}-\omega_{p_2})$ upon time integration \cite{quesada2022beyond}. As a consequence, the frequency domain description of the generated quantum state is naturally provided by the continuum of asymptotic states.  On the contrary, in the high gain regime, it is more practical to solve the system dynamics entirely in the time domain \cite{quesada2022beyond,vernon2015strongly,vernon2019scalable}. We use this approach to write the master equation for the signal/idler density matrix. Following \cite{quesada2022beyond}, we write the full hamiltonian of the resonator coupled to the bus waveguide and the \emph{phantom} loss channel as $H=H_{\textrm{wg}}+H_{\textrm{ph}}+H_{\textrm{r}}+H_{\textrm{c}}+H_{\textrm{NL}}$, where the first three terms describe the free-evolution of the waveguide-phantom channels $a_{\textrm{wg-ph}}(k)$ (forming a continuum) and of the discrete resonator modes $a_{s(i)}$. The cavity-channel coupling is given by
\begin{equation}
    H_c = \hbar\sqrt{v_g}\sum_{J=(i,s)}\left (\sqrt{2\gamma_{eJ}}\Psi_{\textrm{wg,J}}(0)+\sqrt{2\gamma_i}\Psi_{\textrm{ph,J}}(0)\right )a_J^{\dagger}  + \textrm{H.c.},
\label{eq:coupling_ham}
\end{equation}
where $v_g$ is the group velocity and 
\begin{equation}
    \Psi_{\textrm{wg,ph}}(z) = \int \frac{dk}{\sqrt{2\pi}}a_{\textrm{wg,ph}}(k)e^{i(k-k_J)z}.
    \label{eq:field_definition}
\end{equation}
The integration is performed in the neighborhood of the wavevector $k_{J=(s,i)}$ of the signal(idler) resonance. The nonlinear hamiltonian $H_{\textrm{NL}}$ describes the effects of XPM and SFWM, and is given by
\begin{equation}
    H_{\textrm{NL}} = -[\bar{\Lambda}(\langle a_p(t)\rangle^*)^2a_s^{\dagger}a_i^{\dagger} + \textrm{h.c.}] 
    -2\bar{\Lambda}|\langle a_p(t)\rangle|^2 (a_s^{\dagger}a_s + a_i^{\dagger}a_i) 
\label{eq:nl_hamiltonian}
\end{equation}
where 
\begin{equation}
\bar{\Lambda} = \frac{\hbar\omega_p^2 c n_2}{n_0^2V_{\textrm{eff}}}  
\end{equation}
and $\langle a_p(t) \rangle $ is the solution of the differential equation 
\begin{figure*}[t!]
    \includegraphics[scale = 0.35]{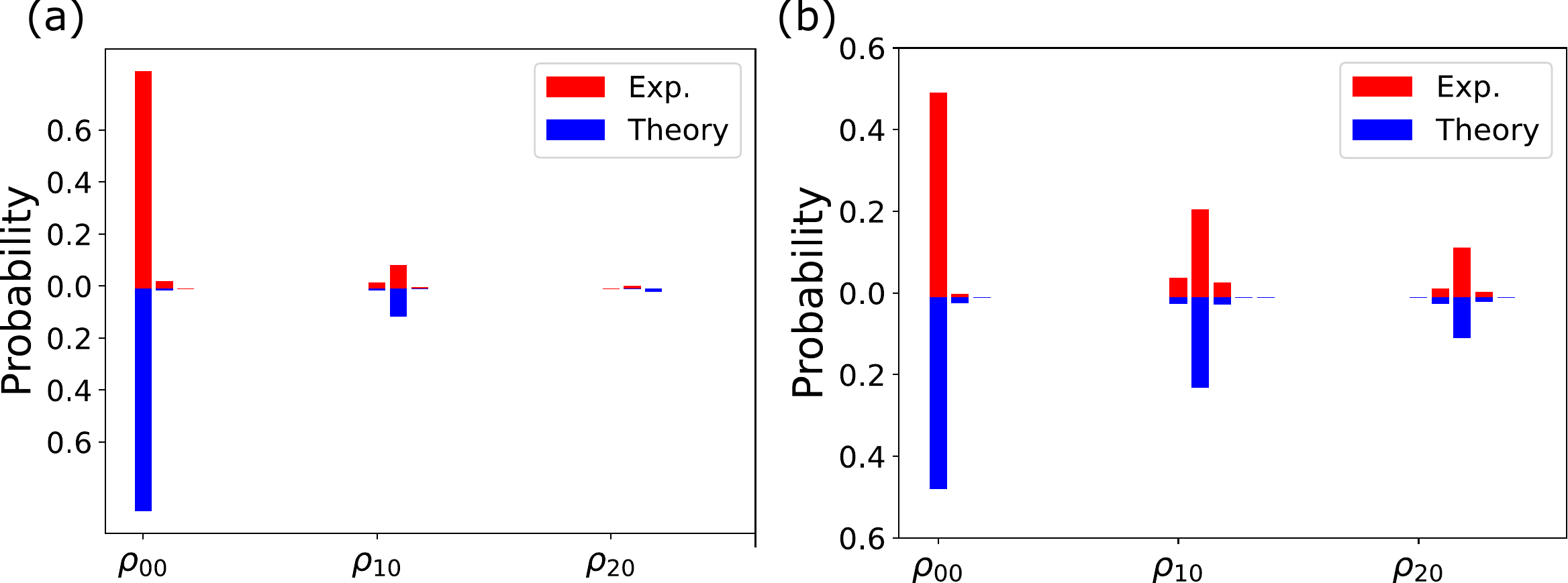}
    \caption{Comparison between the experimental PND of the TMS at the output of the resonator and the one predicted by the stochastic master equation Eq.(\ref{eq:master_eq}). Panels (a) and (b) refer respectively to an input pump power of $1$ mW and $2.5$ mW.} 
    \label{Fig_6}
\end{figure*}
\begin{equation}
\begin{gathered}
    \left[ \frac{d}{dt} +\gamma_{\textrm{tot}}-2i\bar{\Lambda}|\langle a_p \rangle|^2 \right]\langle a_p\rangle = -i\sqrt{2\gamma_{ep}}\beta(t)e^{i\Delta_pt}
\end{gathered}
\label{eq:pump_eq_in_stochastic}
\end{equation}
In writing Eq.(\ref{eq:nl_hamiltonian}) we replaced the pump operators $a_p$ and $a_p^{\dagger}$ with the complex numbers \mbox{$a_p\rightarrow \langle a_p \rangle$} and \mbox{$a_p^{\dagger}\rightarrow \langle a_p\rangle ^*$}, which is the usual mean field approximation for a strong pump in a coherent state \cite{vernon2015strongly}. We also assume that the mean number of pump photons in the cavity $\int |\langle a_p(t) \rangle |^2dt \ll \langle c_{s(i)}^{\dagger}c_{s(i)}\rangle$, which justifies the absence of pump depletion terms in Eq.(\ref{eq:pump_eq_in_stochastic}). Within these approximations, the pump equation is self-contained. We choose the driving field $\beta(t)$ to be a top-hat function of duration \mbox{$T$}
\begin{equation}
    \beta(t) = \sqrt{\frac{P}{R\hbar\omega_p}}\left (\Theta(t)-\Theta(t-T) \right), 
\end{equation}
where $P$ is the average power and $\Theta(t)$ is a step function defined as $\Theta(t)=1$ for $(t>=0)$ and $\Theta(t)=0$ for $(t<0)$. \\
\begin{table}[h!]
\centering
\begin{tabular}{cc}
Parameter & Value \\
\hline \hline
$\gamma_{\textrm{tot,i}}$ & $2.25\times 10^{-3}\,\textrm{ps}^{-1}$ \\ 
$\gamma_{\textrm{tot,p}}$ & $1.84\times 10^{-3}\,\textrm{ps}^{-1}$ \\ 
$\gamma_{\textrm{tot,s}}$ & $2.25\times 10^{-3}\,\textrm{ps}^{-1}$ \\ 
$\eta_{es}$ & $0.926$ \\
$\eta_{ei}$ & $0.926$ \\
$\tau_{\textrm{rt}} $ & $5$ ps\\
$n_g$ & $2.09$ \\
$\bar{\Lambda}$ & $1.72\,\textrm{s}^{-1}$  \\
$\Lambda$ & $6.72\times 10^7\,\textrm{J}^{-1}$  \\

\end{tabular}
\caption{\label{Table_1} List of the parameters used in the simulation of the master equation of the signal and idler modes.}
\end{table}
The master equation for the signal and idler density matrix is obtained by tracing out the continuum of modes of the waveguide and the phantom channel. Due to the colorless model of the minimal coupling between the ring and the channel modes in Eq.(\ref{eq:coupling_ham}), we can use the standard second-order Born-Markov approximation to write the master equation for the signal and the idler modes $a_{s(i)}$ in Lindblad form
\begin{equation}
    \dot{\rho} = -\frac{i}{\hbar}[H_{\textrm{NL}},\rho] + \left ( \mathcal{D}[\sqrt{2\gamma_{\textrm{tot,s}}}a_s]+\mathcal{D}[\sqrt{2\gamma_{\textrm{tot,i}}}a_i] \right )\rho, \label{eq:master_eq}
\end{equation}
where $\mathcal{D}[a] = a\rho a^{\dagger}-\frac{1}{2}(a^{\dagger} a \rho + \rho a^{\dagger} a)$. We numerically solve Eq.(\ref{eq:master_eq}) using the open access Python library Qutip \cite{johansson2012qutip}. The Fock space dimension for the signal and the idler photon is truncated to $12$ excitations, without any noticeable change as a result of moving into a larger space. The list of simulation parameters is shown in Table \ref{Table_1}. The density matrix $\rho(t)$ is used to calculate the average photon number in the cavity $\textrm{Tr}(\rho(t)a_{s(i)}^{\dagger}a_{s(i)})$, which is related to the number of scattered photons in the channel waveguide per pulse $\langle n_{s(i)} \rangle$ as
\begin{equation}
\langle n_{s(i)} \rangle = 2\gamma_{es(i)} \int \textrm{Tr}(\hat{\rho}(t)a_{s(i)}^{\dagger}a_{s(i)}) dt 
\label{eq:mean_photon_number}
\end{equation}
The Schmidt number $K$ can be inferred from the time-integrated $\bar{g}_2$
\begin{equation}
\bar{g}_2 = \frac{\int \langle a^{\dagger}(t)a^{\dagger}(t+\tau)a(t+\tau)a(t)\rangle dt d\tau}{\left ( \int \langle a^{\dagger}(t)a(t)\rangle dt \right )^2}    
\end{equation}
as $K = \bar{g}_2-1$ \cite{christ2011probing}. We calculated the numerator using the Qutip method \verb|qutip.correlation_3op_2t|, which calculates the two-time correlation function by applying the quantum regression theorem, while the denominator is calcuilated using Eq.(\ref{eq:mean_photon_number}).\\ 
The PND is simulated using a stochastic master equation describing the signal and the idler modes subjected to continuous photodetection \cite{wiseman2009quantum}. Since only those photons which are scattered into the waveguide can be detected, we first split the dissipation term in Eq.(\ref{eq:master_eq}) as 
\begin{equation}
\mathcal{D}\left[2\sqrt{\eta_{es(i)}\gamma_{\textrm{tot,s(i)}}}c_{s(i)}\right] + \mathcal{D}\left[\sqrt{2(1-\eta_{es(i)})\gamma_{\textrm{tot,s(i)}}}c_{s(i)}\right] \label{eq:split_D}
\end{equation}
and let the first and the second term to govern respectively the stochastic and the deterministic part of the evolution of each quantum trajectory \cite{wiseman2009quantum,onodera2022nonlinear}. The stochastic master equation is solved by using the method \verb|qutip.stochastic.photocurrent_mesolve| \cite{johansson2012qutip}. To calculate the PND, we simulated $N_\textrm{tot} = 3000$ trajectories and for each of them we recorded the number of joint signal/idler detections $N_{n_s,n_i}$. The dataset is then binned on a two-dimensional grid representing the signal and the idler photon numbers $n_s$ and $n_i$, and the probabilities are approximated as $p(n_s,n_i)\sim N_{n_s,n_i}/N_{\textrm{tot}}$. Figure \ref{Fig_6} shows some examples of simulated PND at different pump powers compared to experimental ones. The fidelities between the simulation and the experiment are reported in Table 2.  
\begin{table}[h!]
\centering
\begin{tabular}{cc}
Pump power (mW) & $\mathcal{F}$ \\

\hline \hline
$1.00$ & $0.985$ \\ 
$1.26$ & $0.984$ \\ 
$1.41$ & $0.986$ \\ 
$1.58$ & $0.982$ \\ 
$1.78$ & $0.981$ \\ 
$2.00$ & $0.978$ \\ 
$2.23$ & $0.976$ \\ 
$2.51$ & $0.956$ \\ 
$2.82$ & $0.941.$ \\ 
\end{tabular}
\caption{\label{Table_2} Fidelity between the simulated PND and the experimental ones for different pump powers.}
\end{table}

%

\end{document}